\documentclass[a4paper,11pt]{article}
\usepackage{graphicx}
\usepackage{fullpage}
\pdfoutput=1 % if your are submitting a pdflatex (i.e. if you have
\usepackage{jheppub} % for details on the use of the package, please
\usepackage{booktabs}
\usepackage{subcaption}
\usepackage{framed}
\usepackage{multirow}
\usepackage{array}
\usepackage{lscape}
\usepackage{slashed}

\title{\sffamily Enhancing $t\bar{t}hh$ production through CP-violating top-Higgs interaction at the LHC and future colliders}

\author[a,b]{Ning Liu,}%\note{Corresponding author: Ning Liu}}
\author[a]{Yanming Zhang,}
\author[c]{Jinzhong Han}
\author[a]{and Bingfang Yang}

\affiliation[a]{Institute of Theoretical Physics, Henan Normal University, Xinxiang 453007, China}
\affiliation[b]{ARC Centre of Excellence for Particle Physics at the Terascale, School of Physics, The University of Sydney, NSW 2006, Australia}
\affiliation[c]{School of Physics and Electromechnical Engineering, Zhoukou Normal University, Zhoukou, 466001, China}

\emailAdd{wlln@mail.ustc.edu.cn}
\emailAdd{zhangyanming@henannu.edu.cn}
\emailAdd{hanjinzhong@zknu.edu.cn}
\emailAdd{yangbingfang@htu.edu.cn}

\abstract{
The measurement of Higgs self-coupling is one of the most crucial physics goals at the future colliders. At the LHC, the di-Higgs production is a main way to measure the Higgs trilinear coupling. As a complementary to the di-Higgs production, $t \bar{t} hh$ process may open a new avenue to measure di-Higgs physics at the LHC and a future 100 TeV $pp$ collider or a high energy $e^+e^-$ collider since the extra $t\bar t$ in the final states may efficiently suppress the backgrounds. However, such a kind of process is also controlled by the top-Higgs coupling. In this work, we investigate the impact of CP-violating top-Higgs coupling on $t\bar{t}hh$ production at the LHC, $e^+e^-$ and a 100 TeV hadron collider under the current Higgs data. Within 2$\sigma$ Higgs data allowed parameter region, we find that the cross section of $t\bar{t}hh$ at the LHC-14 TeV, $e^+e^-$-1 TeV and VHE-LHC/SPPC-100 TeV can be enhanced up to 2.1 times the SM predictions. The future precise measurement of Higgs coupling will reveal the nature of top-Higgs interaction and improve the sensitivity of the determination of Higgs self-coupling through $t\bar{t}hh$ production.}

\begin{document}
\maketitle

\section{Introduction}

In 2012, a bosonic resonance with a mass around 125 GeV was independently observed by the ATLAS and CMS collaborations at the LHC \cite{2012atlas,2012cms}. Up to now, most measurements of its properties are consistent with the predictions of the Standard Model (SM) Higgs boson \cite{2013atlas,2013cms}. However, the Higgs couplings with top quarks and with itself are still vacant and need to be verified at the future colliders.

The tree-level Higgs potential of the SM is given by,
\begin{equation}
V=-\mu^2 (\phi^\dagger\phi)+\lambda(\phi^\dagger\phi)^2,\quad \phi=\frac{1}{\sqrt{2}}\left(0, v+h \right)^T,\label{potential}
\end{equation}
which yields the trilinear and quartic Higgs self-couplings,
\begin{equation}
\lambda_{3h}=\frac{3m^2_{h}}{v}, \quad \lambda_{4h}=\frac{3m^2_{h}}{v^2}\label{3h-4h}.
\end{equation}
where the Higgs boson mass $m_h\simeq 125$ GeV have been well measured in the experiments and the vacuum expectation value of the Higgs field, $v=(\sqrt{2}G_F)^{-1/2}\simeq 246$ GeV. Hence, the determination of the Higgs self-coupling can directly test the relation Eq.(\ref{3h-4h}) obtained from the minimization of SM Higgs potential Eq.(\ref{potential}). At the LHC, the main way to measure the Higgs self-coupling is the di-Higgs production, which is dominated by the gluon-gluon fusion mechanism and has small cross section \cite{hh-sm,Frederix:2014hta,wengan}. Among various decay channels, the process $hh \to b\bar{b}\gamma\gamma$ is expected to have the most promising sensitivity due to the low backgrounds at the LHC \cite{hh-mc1} while the $4b$ final state has the largest fraction \cite{hh-mc2}. Some applications of advanced analysis techniques to di-Higgs production have been proposed to improve the sensitivity of $b\bar{b}\tau^+\tau^-$ and $b\bar{b}W^+W^-$ channel \cite{hh-mc3}. However, one should keep in mind that, the main production process of di-Higgs $gg \to hh$ can also be triggered by the top-Higgs coupling itself through the box diagrams. So, any new physics in the top-Higgs coupling may significantly affect the measurement of Higgs self-coupling at the LHC \cite{liuning,lilin,liantao,contino,susy,nishiwaki,ian,das}.

In the SM, top quark has the strongest coupling to the Higgs boson and is widely speculated as a sensitive probe to the new physics beyond the SM. The most sensitive direct way of measuring the top-Higgs coupling at the LHC is the associated production of the top pair with Higgs boson that has been extensively studied in literature \cite{tth-atlas,tth-cms,tth-old,tth-new}. However, since the cross section of $pp \to t\bar{t} h$ is about 130 fb at 8 TeV LHC \cite{Heinemeyer:2013tqa}), current LHC luminosity and analysis are not yet sensitivity enough to observe such a signal. An upper limit on the signal strength $\mu_{t\bar{t}h}=\sigma_{t\bar{t}h}/\sigma^{SM}_{t\bar{t}h}$ to be $\mu_{t\overline{t}h}\lesssim3.9$ at 95\% C.L. limit has been set up by ATLAS collaboration through combining $h \to b \bar{b}$ and $h \to \gamma\gamma$ channels \cite{atlas:tth_gaga}. While the CMS collaboration gives a limit of $0.9 \lesssim \mu_{t\bar{t}h} \lesssim 3.5$ by using all search channels \cite{Khachatryan:2014qaa}. The other direct way is to observe the single top associated production with the Higgs boson. Such a process has a small cross section (18.28 fb at 8 TeV LHC) but is advocated to determine the sign of the top-Higgs coupling at the LHC \cite{thj-old,thj-new}. Recently, the CMS collaboration has presented the result on $thj$ searches in the $h \to \gamma\gamma$ channel, and obtained a weak bound on the cross section of events with inverted top-Higgs coupling \cite{thj-cms}.

In this work, we will investigate the effect of non-standard top-Higgs coupling in the $t\bar{t}hh$ production at the LHC, a $e^+e^-$ and a 100 TeV hadron collider under the current Higgs data constraints. At the LHC, although the cross section of $pp \to t\bar{t}hh$ is about an order smaller than that of $pp \to hh$, the additional $t\bar t$ in the final states may suppress one order or orders more backgrounds \cite{tthh1,tthh2}. Besides, such a process has a cross section monotonically increasing with respect to Higgs self-coupling \cite{Frederix:2014hta}, which may complement the process $pp \to hh$ in measuring Higgs self-coupling, particularly for $\lambda \gg \lambda_{SM}$. So, the process $pp \to t \bar t hh$ opens a new avenue to measure di-Higgs physics at HL-LHC and a future 100 TeV $pp$-collider. On the other hand, given the limited precision of the LHC, an $e^+e^-$ collider is crucial to scrutinize the detailed properties of the Higgs boson that might uncover the new physics beyond the SM \cite{ilc-white}. In $e^+e^-$ collision, the double Higgs strahlung $e^+e^- \to Zhh$ is considered as the main production process to measure the Higgs self-coupling. Besides, it should be noted that the di-Higgs bosons can also be radiated off from top quarks through the process $e^+e^- \to t\bar{t}hh$, or $\gamma\gamma \to t\bar{t}hh$ at photon collider, where the energetic electron beam to a photon beam is converted through the backward Compton scattering \cite{ilc-rr}. Similar to $pp \to t\bar{t}hh$, the processes $e^+e^-/\gamma\gamma \to t\bar{t}hh$ not only involve the Higgs self-coupling but also are governed by the top-Higgs coupling. So the investigation of $t\bar{t}hh$ production may lead us to obtain the nontrivial information on the Higgs potential and test the top-Higgs coupling \cite{ee-tthh,aa-tthh}.

The structure of this paper is organized as follows. In Section \ref{section2}, we will briefly introduce the non-standard top-Higgs interaction and set up the calculations. In Section \ref{section3}, we present the numerical results and discuss the effects of non-standard top-Higgs coupling in the di-Higgs production at the LHC, $e^+e^-$ and a 100 TeV hadron collider. Finally, we draw our conclusions in Section \ref{section4}.

\section{CP-violating Top-Higgs Couplings}\label{section2}

In this study, we parameterize the top-Higgs couplings using the phenomenological Lagrangian:
\begin{equation}
{\cal L} = -\frac{y_t}{\sqrt{2}}\overline{t}(\cos\theta+i\gamma^{5}\sin\theta)th.
\label{lag}
\end{equation}
In the SM, $y_t$ takes the value $y^{SM}_t=\sqrt{2}m_t/v$ and $\sin\theta =0$, with $v \simeq 246$ GeV being the vacuum expectation value of the Higgs field. A pure pseudo-scalar interaction can be obtained by setting $\cos\theta=0$. A CP violating interaction is realized if both $\cos\theta\neq0$ and $\sin\theta\neq0$. The exact values of these coefficients depend on the specific model. Here we are interested in a model-independent approach to determine the impact of general top-Higgs coupling on $t\bar{t}hh$ production. In the following calculations, we define two reduced couplings: $c_t= y_t \cos\theta/y_{t_{SM}}$ and $\tilde{c}_t= y_t \sin\theta/y_{t_{SM}}$ to discuss our results.

Since the CP-violating top-Higgs interaction can sizably alter the production rate of $gg \to h$ and decay width of $h \to \gamma\gamma$ through the loop effect, the most relevant indirect constraint on the couplings $c_t$ and $\tilde{c}_t$ should be from the Higgs data. The signal strength of one specific analysis from a single Higgs boson is given by
\begin{equation}
\mu = \sum_{i} \mu_i\omega_i,
\label{Eq:mu}
\end{equation}
where the sum runs over all channels used in the experimental analysis. Each channel is characterized by one specific production and decay mode. The
individual channel signal strength can be calculated by
\begin{equation}
\mu_i=\frac{\left[\sigma\times BR\right]_i}{\left[\sigma_{SM}\times BR_{SM}\right]_i},
\label{Eq:ci}
\end{equation}
and the SM channel weight is
\begin{equation}
\omega_i=\frac{\epsilon_i\left[\sigma_{SM}\times BR_{SM}\right]_i}{\sum_j\epsilon_j\left[\sigma_{SM}\times BR_{SM}\right]_j}.
\label{Eq:omega}
\end{equation}
where $\epsilon_i$ is the relative experimental efficiencies for each channel. But these are rarely quoted in experimental publications. In this case, all channels considered in the analysis are treated equally, i.e. $\epsilon_i=1$. The reduced effective coupling $c_{hgg}$ and $c_{h\gamma\gamma}$ can be parameterized through the $c_t$ and $\tilde{c}_t$ as following \cite{zupan},
\begin{eqnarray}
c^2_{hgg} \; & \simeq & c_t^2 + 2.6 {\tilde c}_t^2 + 0.11 c_t (c_t - 1) \, , \nonumber \\
c^2_{h\gamma\gamma} \; & \simeq & (1.28 - 0.28 c_t)^2 + (0.43 {\tilde c}_t )^2 \, .
\label{coulping}
\end{eqnarray}
We confront the effective coupling $c_{hgg}$ and $c_{h\gamma\gamma}$ with the Higgs data by calculating the $\chi^{2}$ of the Higgs sector with the public package \textsf{HiggsSignals-1.3.2} \cite{higgssignals}. We choose the mass-centered $\chi^2$ method in the package \textsf{HiggsSignals}. Although the CP-violating top-Higgs interaction can contribute to the electric dipole moment (EDM), the bounds on the coupling $\tilde{c}_t$ strongly rely on the assumption of Higgs couplings to other light fermions \cite{top_cp_review}. But such light quark Yukawa couplings are generally unobservable at the LHC, we do not impose EDM constraints in the study. Other low-energy physics constraints, such as $B_s-\bar{B}_s$ and $B \to X_s \gamma$, are still too weak due to the large uncertainty \cite{zupan}.

\section{Numerical Resutls and Discussions}\label{section3}
In our numerical calculations, we take the input parameters of the SM as \cite{pdg}
\begin{eqnarray}
m_t=173.07{\rm ~GeV}, ~~m_W = 80.385~, ~~m_{Z}=91.19 {\rm
~GeV},\nonumber
\\m_h=125.9{\rm ~GeV},~~\sin^{2}\theta_W=0.2228, ~~\alpha(m_Z)^{-1}=127.918.
\end{eqnarray}
For the strong coupling constant $\alpha_s(\mu)$, we use its 2-loop evolution with QCD parameter $\Lambda^{n_{f}=5}=226{\rm ~MeV}$ and get $\alpha_s(m_Z)=0.118$. We use CT10 parton distribution functions (PDF) for the calculation of $pp \to t\bar{t}hh$ \cite{cteq}. The renormalization scale $\mu_R$ and factorization scale $\mu_F$ are chosen to be $\mu_R=\mu_F=(m_h+m_t)$. It should be noted that the higher order corrections are usually needed to improve the reliability of the leading order results. However, such calculations for four-body production involve complicated techniques and are beyond our study in this work. In the following we will use the ratios of the leading order cross sections $\sigma_{t\bar{t}hh}/\sigma^{SM}_{t\bar{t}hh}$ to present our results, which have the weak dependence on the variation of the scale. Besides, it is expected that the high order effect on the cross sections can be largely canceled in the ratios. We perform the numerical calculations by the package \textsf{calchep-3.4} \cite{calchep}.

\subsection{LHC and VHE-LHC/SPPC}
%%Fig.1 %%%%%%%%%%%%%%%%%%%%
\begin{figure}[ht]
\centering
\includegraphics[width=6in,height=4.1in]{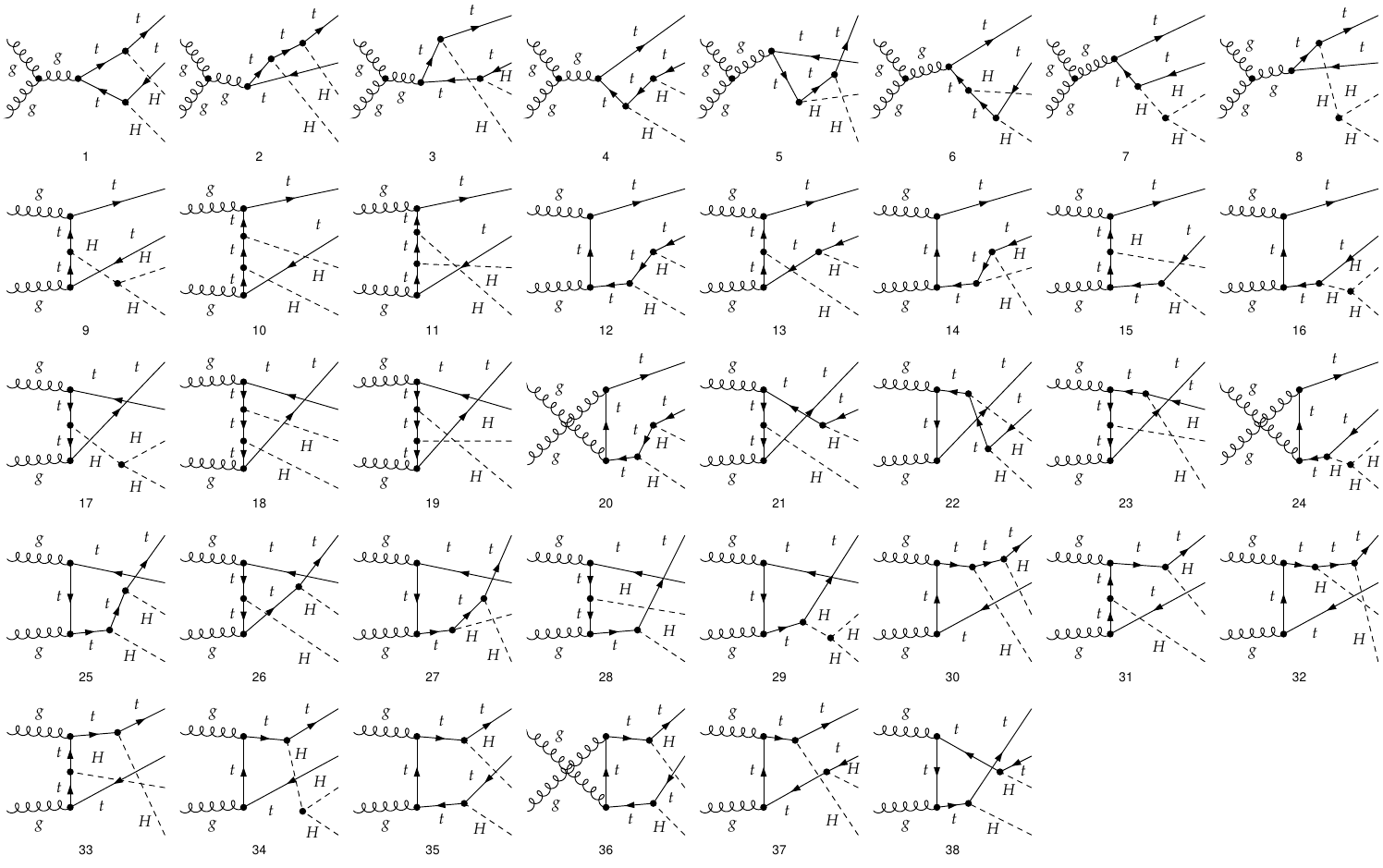}
\caption{Feynman diagrams for the partonic process $gg \to t\bar{t}hh$ at hadron collider.}
\label{gg}
\end{figure}
%%%%%%%%%%%%%%%%%%%%%%%%%
At hadron collider, the main contributions to $t\bar{t}hh$ production are from the gluon fusion processes. The corresponding Feynman diagrams for the partonic process $gg \to t\bar{t}hh$ are shown in Fig.~\ref{gg}. The $q\bar{q}$ annihilation processes can be obtained by replacing the initial gluons with $q\bar{q}$ in the $s$-channel in Fig.~\ref{gg}. According to the production of di-Higgs bosons, we can classify amplitudes of $pp \to t\bar{t}hh$ into two categories: one is proportional to $\alpha_s y^2_t$; the other one is proportional to $\alpha_s y_t\lambda_{3h}$. In the SM, both amplitudes have the same sign and are constructive, which lead to the cross section of $pp \to t\bar{t}hh$ monotonically increases with respect to Higgs self-coupling $\lambda_{3h}$. However, it should be noted that each amplitude is modulated by the top-Higgs coupling. So the measurement of Higgs self-coupling through $pp \to t\bar{t}hh$ strongly depends on the determination of top-Higgs coupling.

%%Fig.2 %%%%%%%%%%%%%%%%%%%%
\begin{figure}[ht]
\centering
\includegraphics[width=3in,height=3in]{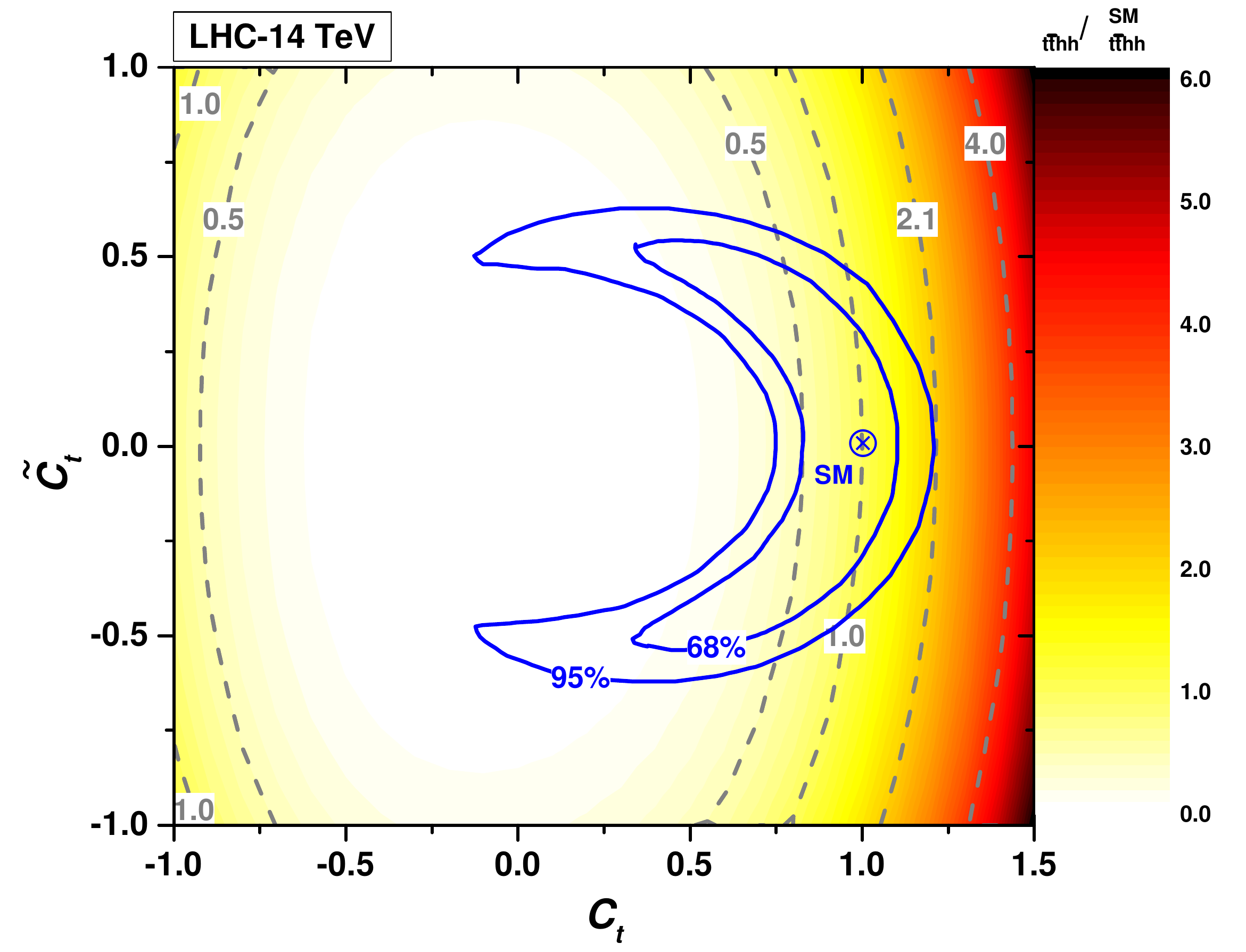}\hspace{-0.3cm}
\includegraphics[width=3in,height=3in]{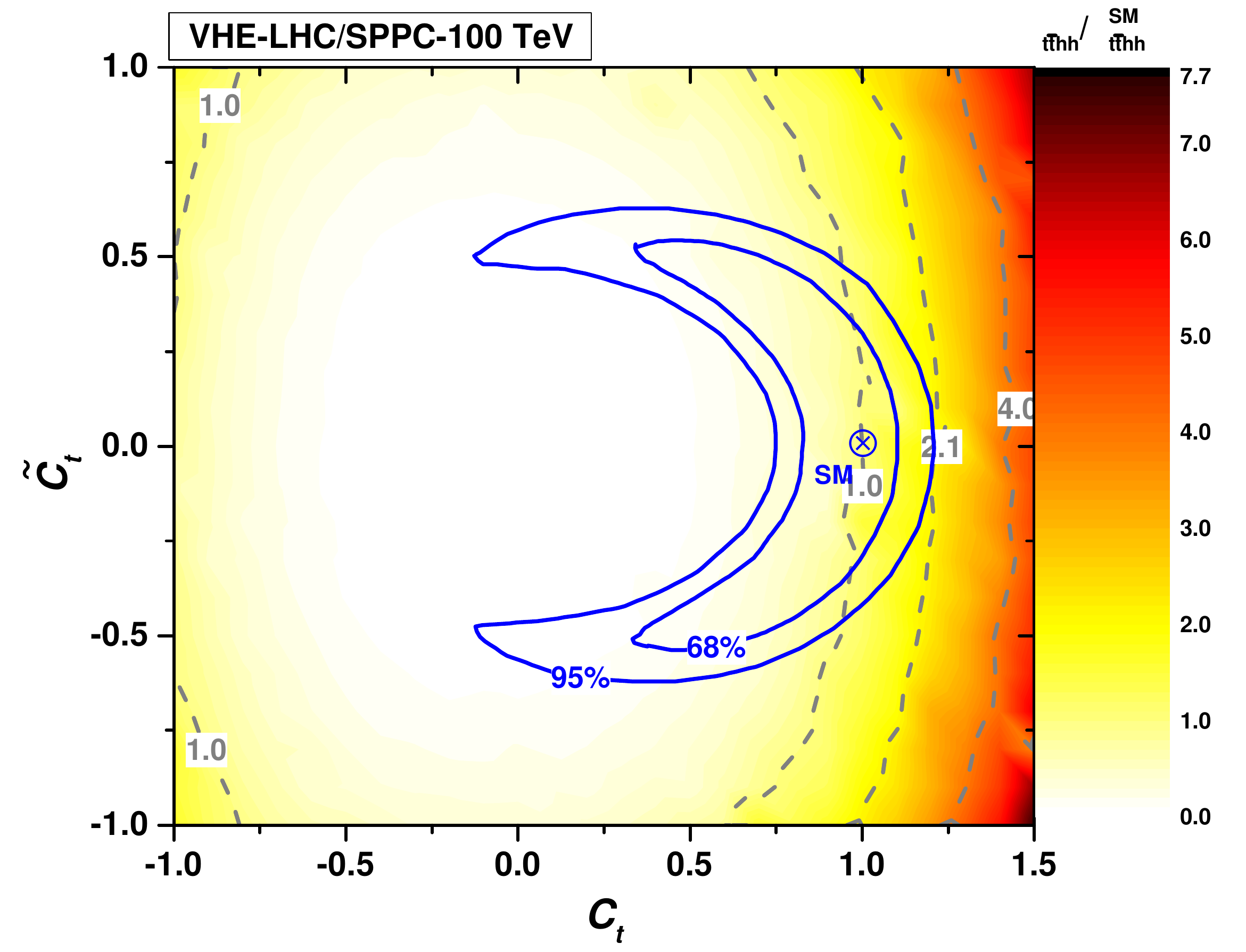}
\caption{Ratios of $\sigma^{pp \to t\bar{t}hh}/\sigma^{pp \to t\bar{t}hh}_{SM}$ in the plane of $\tilde{c}_t-c_t$ at 14 TeV LHC and VHE-LHC/SPPC, where the dashed contours correspond to the 68\% C.L. and 95\% C.L. limits given by the current Higgs data fitting.}
\label{ratio-hadron}
\end{figure}
%%%%%%%%%%%%%%%%%%%%%%%%%
In Fig.~\ref{ratio-hadron}, we respectively show the ratios of $\sigma^{pp \to t\bar{t}hh}/\sigma^{pp \to t\bar{t}hh}_{SM}$ in the plane of $\tilde{c}_t-c_t$ at 14 TeV LHC and VHE-LHC/SPPC, where the dashed contours correspond to the 68\% C.L. and 95\% C.L. limits given by the current Higgs data fitting. From Fig.~\ref{ratio-hadron}, we can see that the negative scalar component $c_t$ in Eq.~\ref{lag} is strongly disfavoured because the experimental measurement of the Higgs diphoton rate are consistent with the SM prediction. The pseudoscalar component in the range $|\tilde{c}_t| \gtrsim 0.6$ have been excluded at 95\% C.L by the Higgs data fitting. Besides, when $c_t=1$ (or $y_t=y^{SM}_t$), a narrow region of pseudoscalar component $|\tilde{c}_t| \lesssim 0.3 (0.44)$ is still allowed at 68\% (95\%) C.L..

As mentioned before, the cross section of $pp \to t\bar{t}hh$ not only depends on the magnitude of the top-Higgs coupling but also on the relative phase angle between between $c_t$ and $\tilde{c}_t$. (i) When $c_t=-1$ and $\tilde{c}_t=0$ (or $y_t=-y^{SM}_t$), the ration of $\sigma^{pp \to t\bar{t}hh}/\sigma^{pp \to t\bar{t}hh}_{SM}$ is smaller than 1 due to the deconstructive interference between the processes with coupling $\alpha_s y^2_t$ and those with coupling $\alpha_s y_t\lambda_{3h}$; (ii) when $c_t=0$ and $\tilde{c}_t=1$ (or $y_t=iy^{SM}_t$), there is no interference between the terms with $\alpha_s y^2_t$ and those with $\alpha_s y_t\lambda_{3h}$, which will reduce the cross section of $pp \to t\bar{t}hh$. Finally, we find that the value of the ratio $\sigma^{pp \to t\bar{t}hh} /\sigma^{pp \to t\bar{t}hh}_{SM}$ can maximally reach about 1.5 (2.1) in the 68\% (95\%) C.L. allowed region at 14 TeV LHC and VHE-LHC/SPPC.

%%Fig.3 %%%%%%%%%%%%%%%%%%%%
\begin{figure}[ht]
\centering
\includegraphics[width=2.5in,height=2.5in]{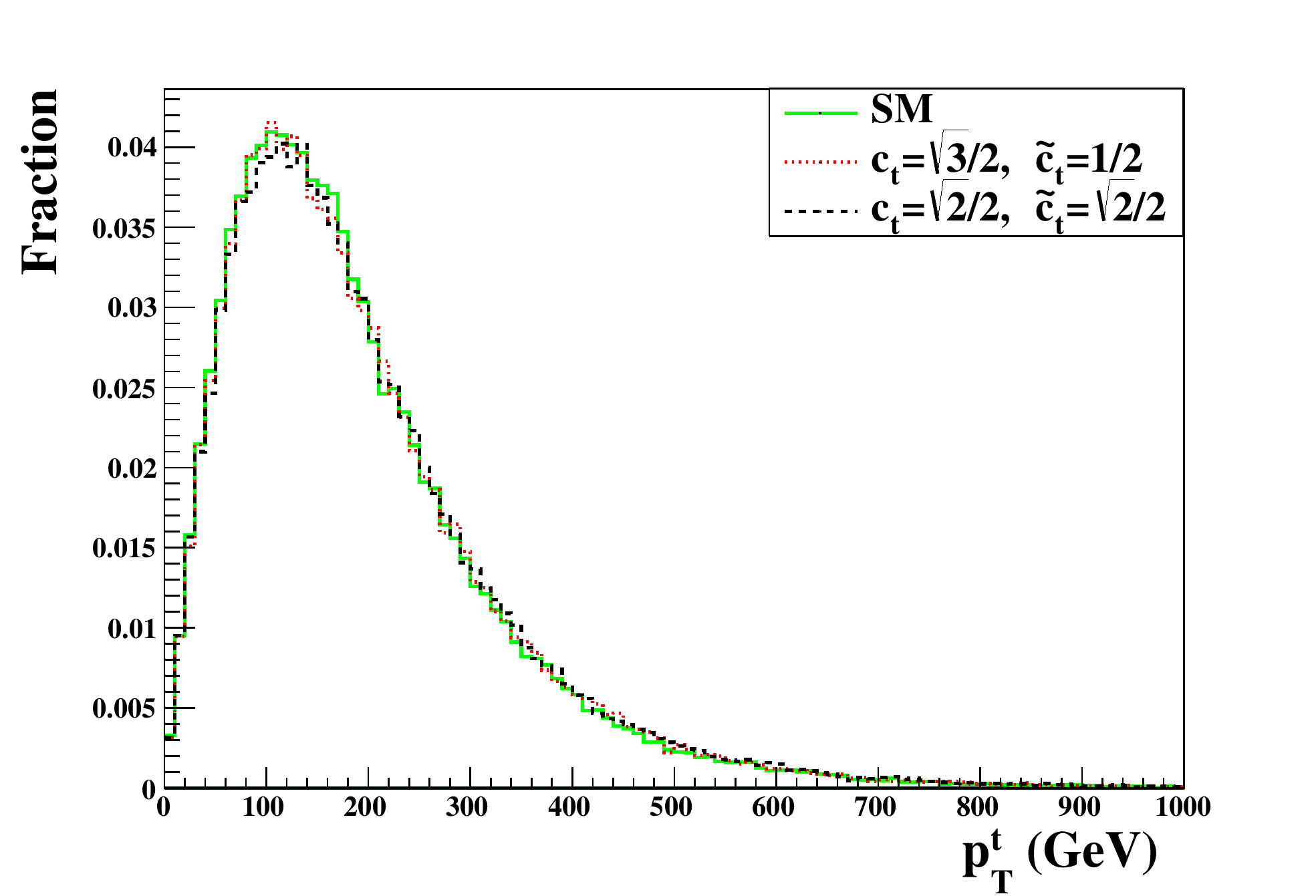}
\includegraphics[width=2.5in,height=2.5in]{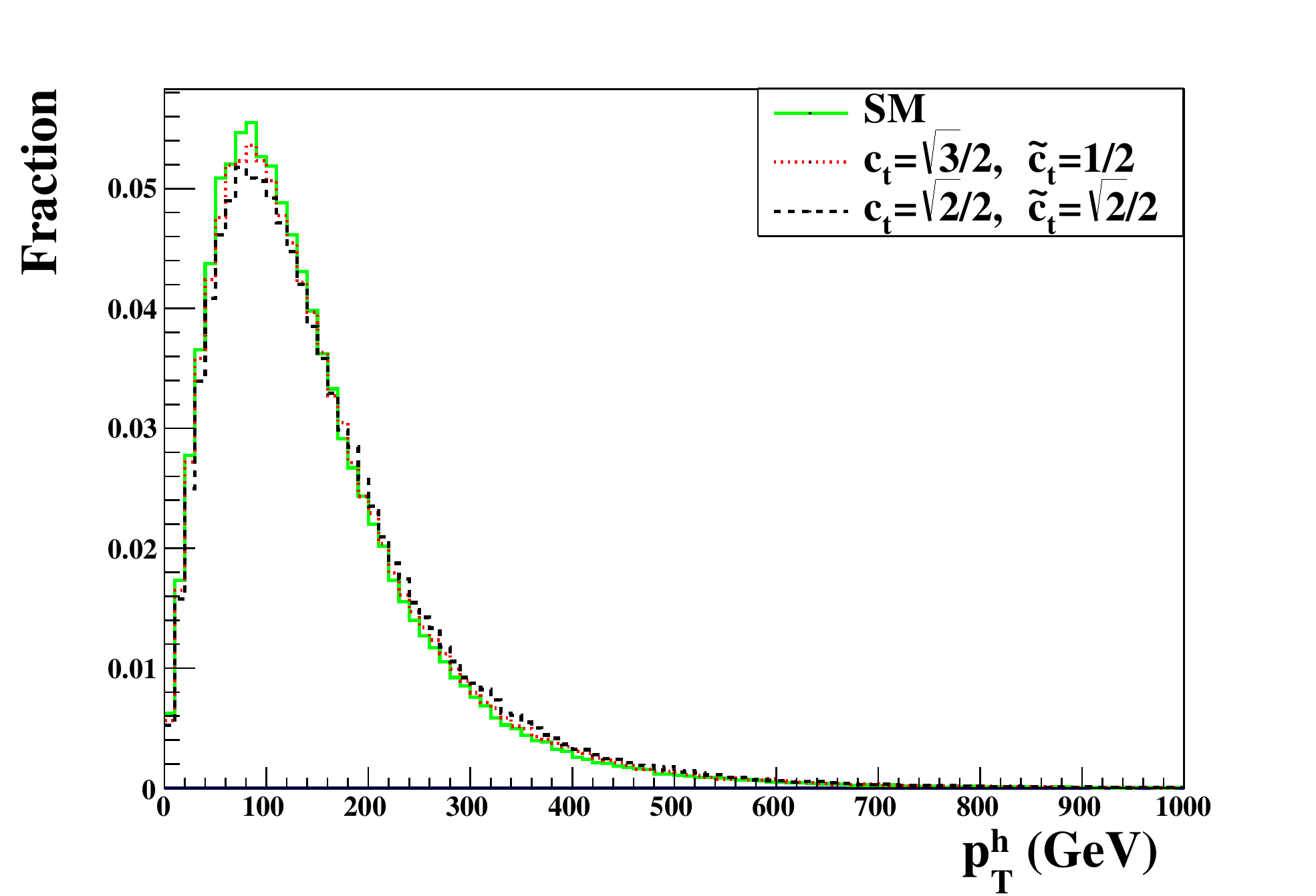}
\caption{The parton-level transverse momentum distributions of top quark and Higgs boson at 14 TeV LHC.}
\label{pt}
\end{figure}
%%%%%%%%%%%%%%%%%%%%%%%%%
We implement the CP-violating interaction of $t\bar{t}h$ in Eq.~\ref{lag} with the package FeynRules \cite{feynrules}. In Fig.~\ref{pt}, we show the parton-level transverse momentum distributions of top quark and Higgs boson in the presence of the CP-violating interactions by using the package \textsf{MadGraph5} \cite{mad5} at 14 TeV LHC. From Fig.~\ref{pt}, we can see that the shapes of $p_T$ distributions of top quark and Higgs boson in the SM are almost the same as those in the CP-violating interactions.

%%Fig.3 %%%%%%%%%%%%%%%%%%%%
\begin{figure}[ht]
\centering
\includegraphics[width=2.5in,height=2.5in]{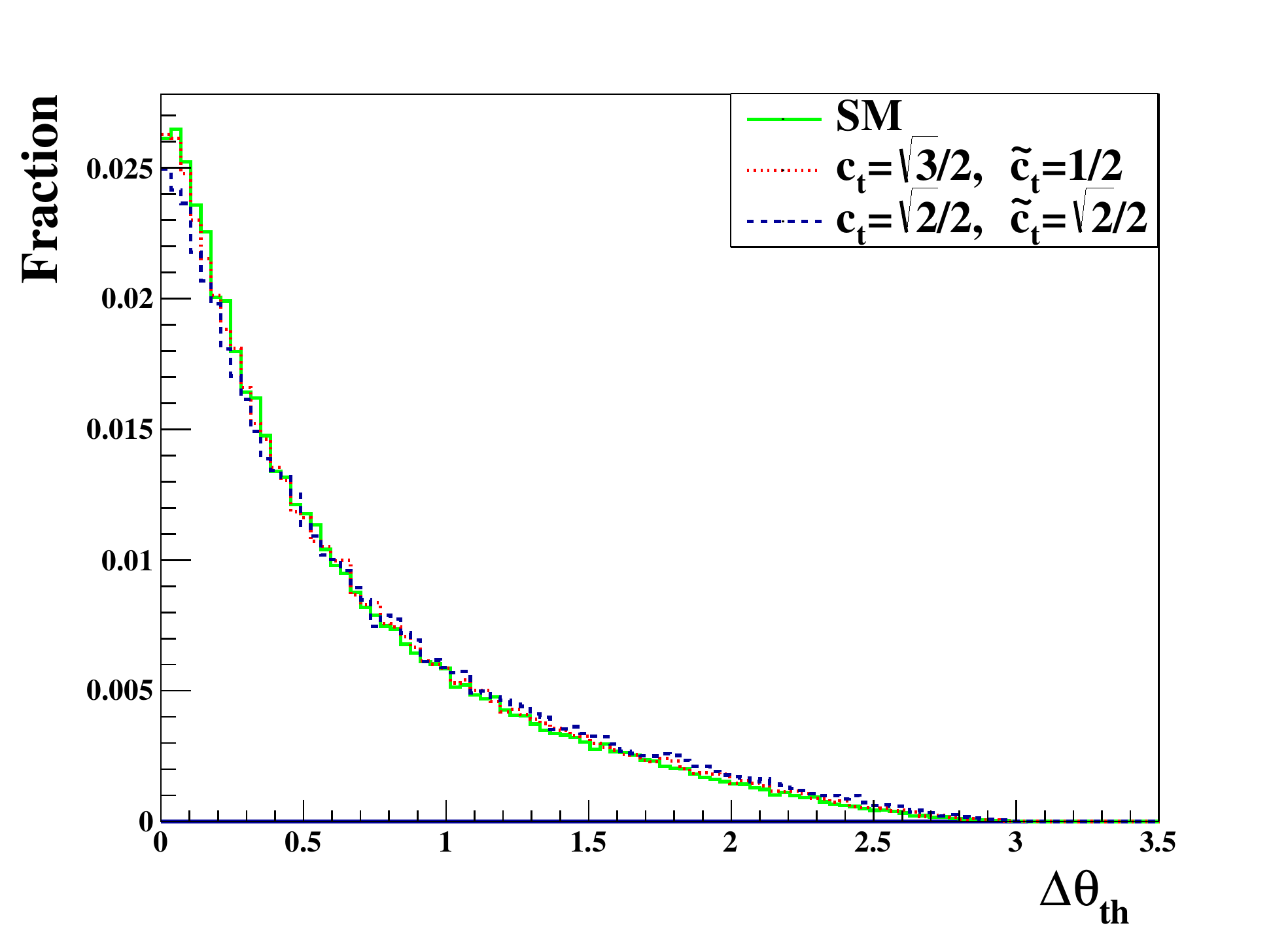}
\includegraphics[width=2.5in,height=2.5in]{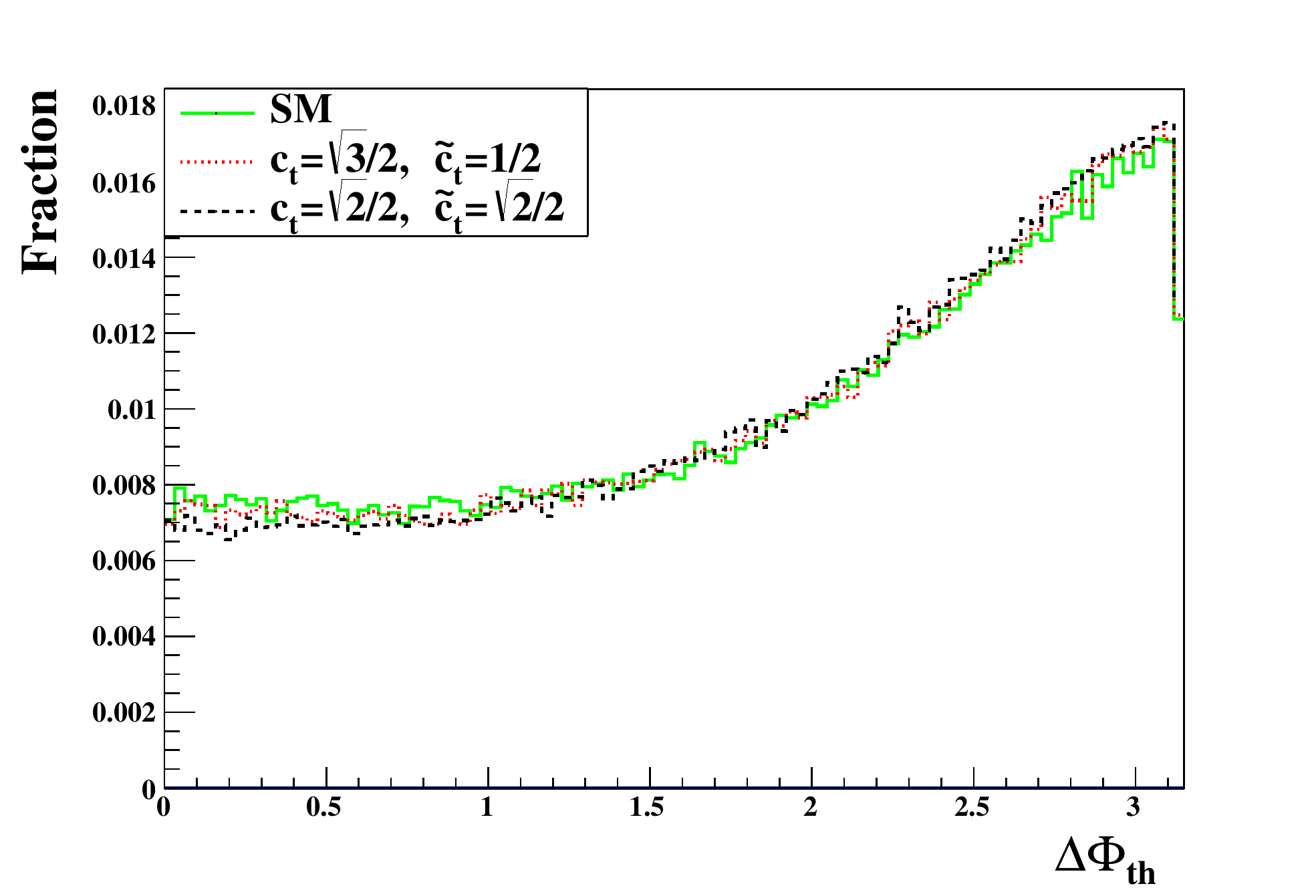}
\caption{The parton-level distributions of polar angle $\Delta\theta_{th}$ and the azimuthal angle $\Delta\phi_{th}$ between the top quark and the Higgs boson at 14 TeV LHC.}
\label{angle}
\end{figure}
%%%%%%%%%%%%%%%%%%%%%%%%%
Since the CP-violating $t\bar{t}h$ interaction in Eq.~\ref{lag} may affect the polarization states of the top quark, we present the parton-level distributions of polar angle $\Delta\theta_{th}$ and the azimuthal angle $\Delta\phi_{th}$ between the top quark and the Higgs boson. From Fig.~\ref{angle}, we can find that the CP-violating interactions slightly change the distributions of $\Delta\theta_{th}$ and $\Delta\phi_{th}$, as the comparison with the SM. To be specific, when $\Delta\theta_{th} \leqslant \pi/2$ ($\Delta\theta_{th} > \pi/2$), the SM has more (less) events than the CP-violating interactions. Similar results can be observed in the distribution of $\Delta\phi_{th}$. Here it should be mentioned that such differences will be more obvious if the pseudo-scalar component $\tilde{c}_t$ becomes large. However, as shown in Fig.~\ref{ratio-hadron}, the current Higgs data has already tightly constrained the size of $\tilde{c}_t$. Besides, these differences will be further diluted when the detector effects are taken into account.

The LHC observability of $t\bar{t}hh$ in the SM including the parton shower and detector effects has been investigated through both Higgs bosons decaying $h \to b\bar{b}$ and semi-leptonic and hadronic top decays in Ref.~\cite{tthh1}. The main backgrounds are $t\bar{t}hb\bar{b}$, $t\bar{t}b\bar{b}b\bar{b}$ and $t\bar{t}Zb\bar{b}$ productions. Due to the four $b$-jets in the signal $t\bar{t}hh$ coming from two Higgs bosons, the most efficient cut is the reconstruction of the Higgs mass through selecting the Higgs decay jets by minimizing, $\chi^2_{HH}=\frac{(m_{b_i,b_j}-m_h)^2}{\Delta^2_h}+\frac{(m_{b_k,b_l}-m_h)^2}{\Delta^2_h}$, where $i\neq j \neq k \neq l$ run over all $b$-tagged jets and $\Delta_h=20$ GeV. To remove the $W+$jets background, at least one top quark is required to be reconstructed by using $\chi^2_{t_l}=\frac{(m_{b_i,l,\slashed E_T}-m_t)^2}{\Delta^2_t}$ for semi-leptonic and $\chi^2_{t_h}=\frac{(m_{j_i,j_j,j_k}-m_t)^2}{\Delta^2_t}$ for hadronic top decays, respectively. Since the Higgs data has tightly constrained the CP-violating $t\bar{t}h$ interactions (see Fig.~\ref{ratio-hadron}), the allowed values of $c_t$ and $\tilde{c}_t$ can not sizably affects the polarization states of the top quark and will not change the distributions of the final states. Therefore, we can expect that the cuts efficiency obtained in the Ref.~\cite{tthh1} is applicable to our case, in particular for pure scalar $t\bar{t}h$ interaction. In this case, we can estimate the LHC sensitivity of our signals by normalizing the results in Ref.~\cite{tthh1}. For example, when $c_t=1.12$ and $\tilde{c}_t=\pm0.23$ (at 95\% C.L. Higgs data allowed region), $\sigma^{pp \to t\bar{t}hh} /\sigma^{pp \to t\bar{t}hh}_{SM}=2$ and $S/B=30\%$. The corresponding statistical significance can reach $3\sigma$ at 14 TeV LHC with the luminosity ${\cal L}=3000$ fb$^{-1}$. So we may expect that the enhancement of $t\bar{t}hh$ arising from the CP-violating $t\bar{t}h$ interactions may be observed at the HL-LHC.

\subsection{$e^+e^-$ and $\gamma\gamma$ collisions}
%%Fig.3 %%%%%%%%%%%%%%%%%%%%
\begin{figure}[ht]
\centering
\includegraphics[width=6in,height=4.5in]{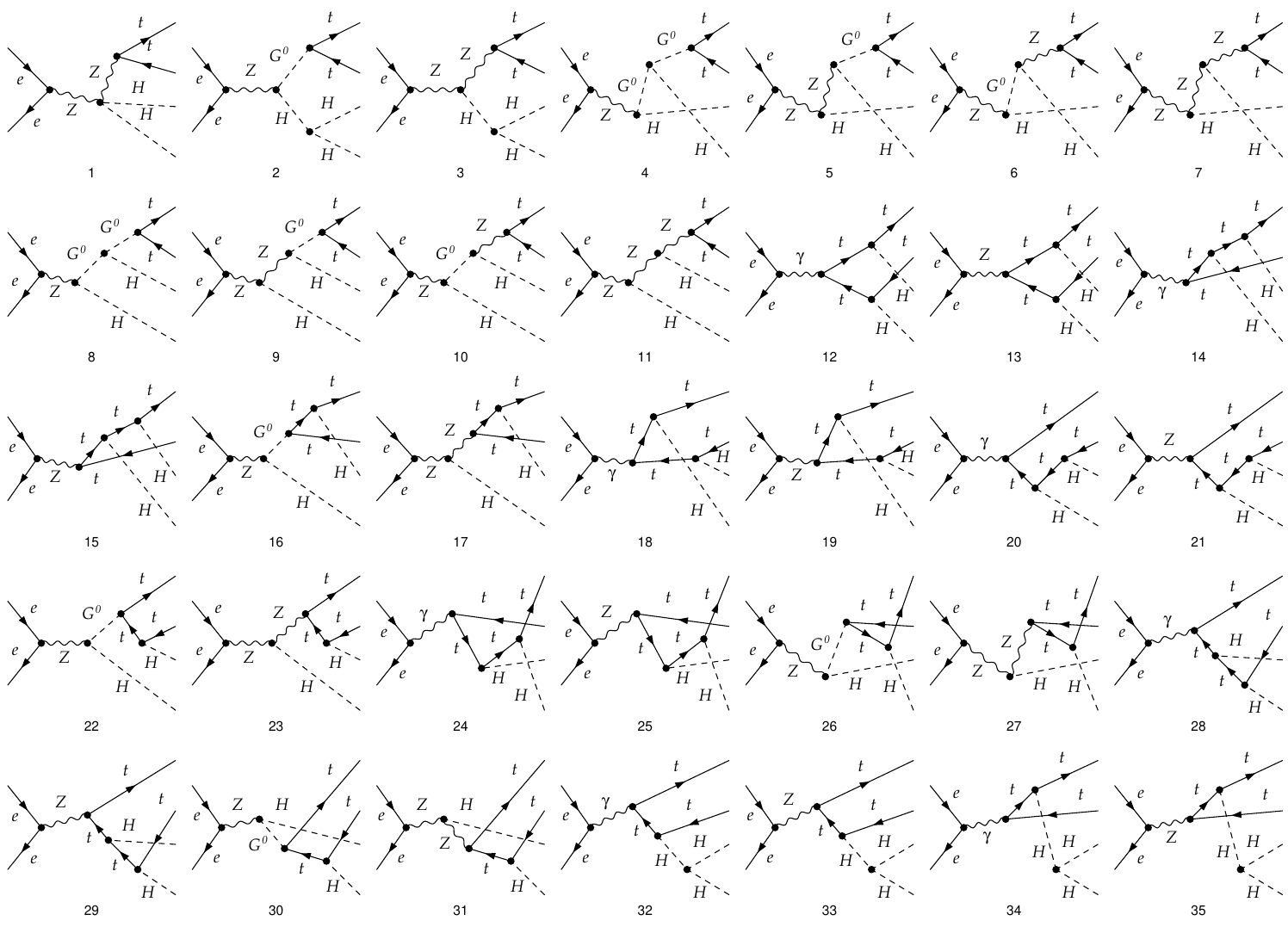}
\caption{Feynman diagrams for the process $e^+e^- \to t\bar{t}hh$.}
\label{ee}
\end{figure}
%%%%%%%%%%%%%%%%%%%%%%%%%
At a high energy $e^+e^-$ collider, $e^+e^- \to t\bar{t}hh$ process can occur only through the $s$-channel. The corresponding Feynman diagrams for $e^+e^- \to t\bar{t}hh$ are shown in Fig.\ref{ee}. Different from the process $pp \to t\bar{t}hh$, the di-Higgs bosons can be produced not only through the top-Higgs or Higgs self interaction but also by the Higgs gauge couplings $hZZ$ or $hhZZ$. Since we assume that the Higgs gauge couplings and Higgs self-couplings be the SM values, any modifications in $t\bar{t}h$ coupling can change the interference behavior between the amplitudes with $\alpha_s y^2_t$ and with $\alpha_s y_t\lambda_{3h}$.

%%Fig.4 %%%%%%%%%%%%%%%%%%%%
\begin{figure}[ht]
\centering
\includegraphics[width=4in,height=3in]{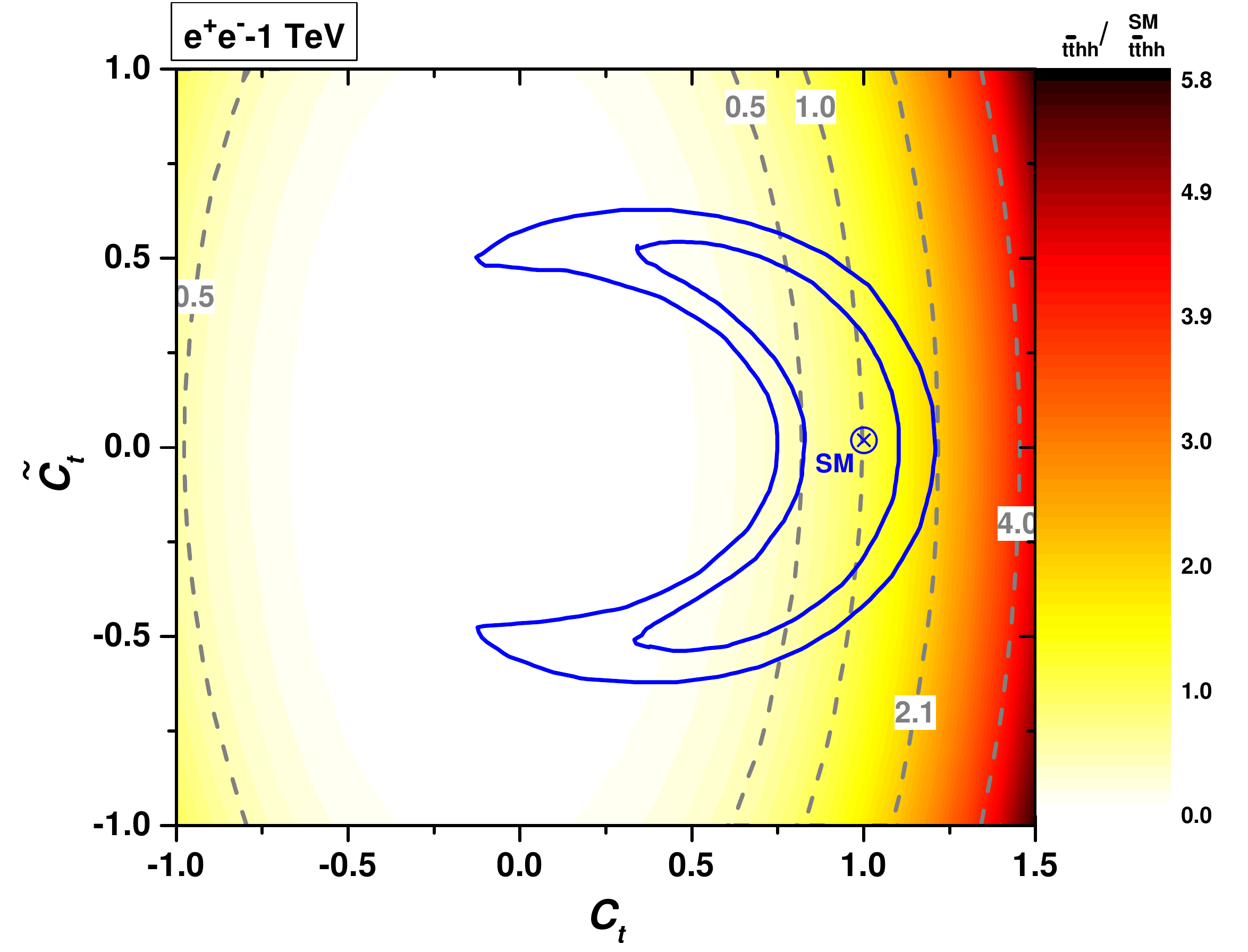}
\caption{Same as Fig.\ref{ratio-hadron}, but ratios of $\sigma^{e^+e^- \to t\bar{t}hh}/\sigma^{e^+e^- \to t\bar{t}hh}_{SM}$ at a $e^+e^-$ collider with $\sqrt{s}=1$ TeV.}
\label{ratio-ee}
\end{figure}
%%%%%%%%%%%%%%%%%%%%%%%%%
In Fig.\ref{ratio-ee}, we show the ratios of $\sigma^{e^+e^- \to t\bar{t}hh}/\sigma^{e^+e^- \to t\bar{t}hh}_{SM}$ in the plane of $\tilde{c}_t-c_t$ at a $e^+e^-$ collider with $\sqrt{s}=1$ TeV. From Fig.\ref{ratio-ee}, we can see that the cross section of $e^+e^- \to t\bar{t}hh$ monotonically increases with $c_t$ when $\tilde{c}_t=0$ and can maximally reach 4 times as large as the SM prediction. For a given $c_t$, the pseudo-scalar component in $t\bar{t}h$ can reduce or enhance the cross section of $e^+e^- \to t\bar{t}hh$. Although some of the subprocesses of $e^+e^- \to t\bar{t}hh$ that can be triggered by the Higgs gauge interactions, the cross section of $e^+e^- \to t\bar{t}hh$ is still dominated by top-Higgs coupling. Within the 68\% (95\%) C.L. Higgs data allowed region, we note that the value of the ratio $\sigma^{e^+e^- \to t\bar{t}hh} /\sigma^{e^+e^- \to t\bar{t}hh}_{SM}$ can maximally reach about 1.5 (2.1) at 1 TeV $e^+e^-$ collider.

%%Fig.3 %%%%%%%%%%%%%%%%%%%%
\begin{figure}[ht]
\centering
\includegraphics[width=6in,height=4in]{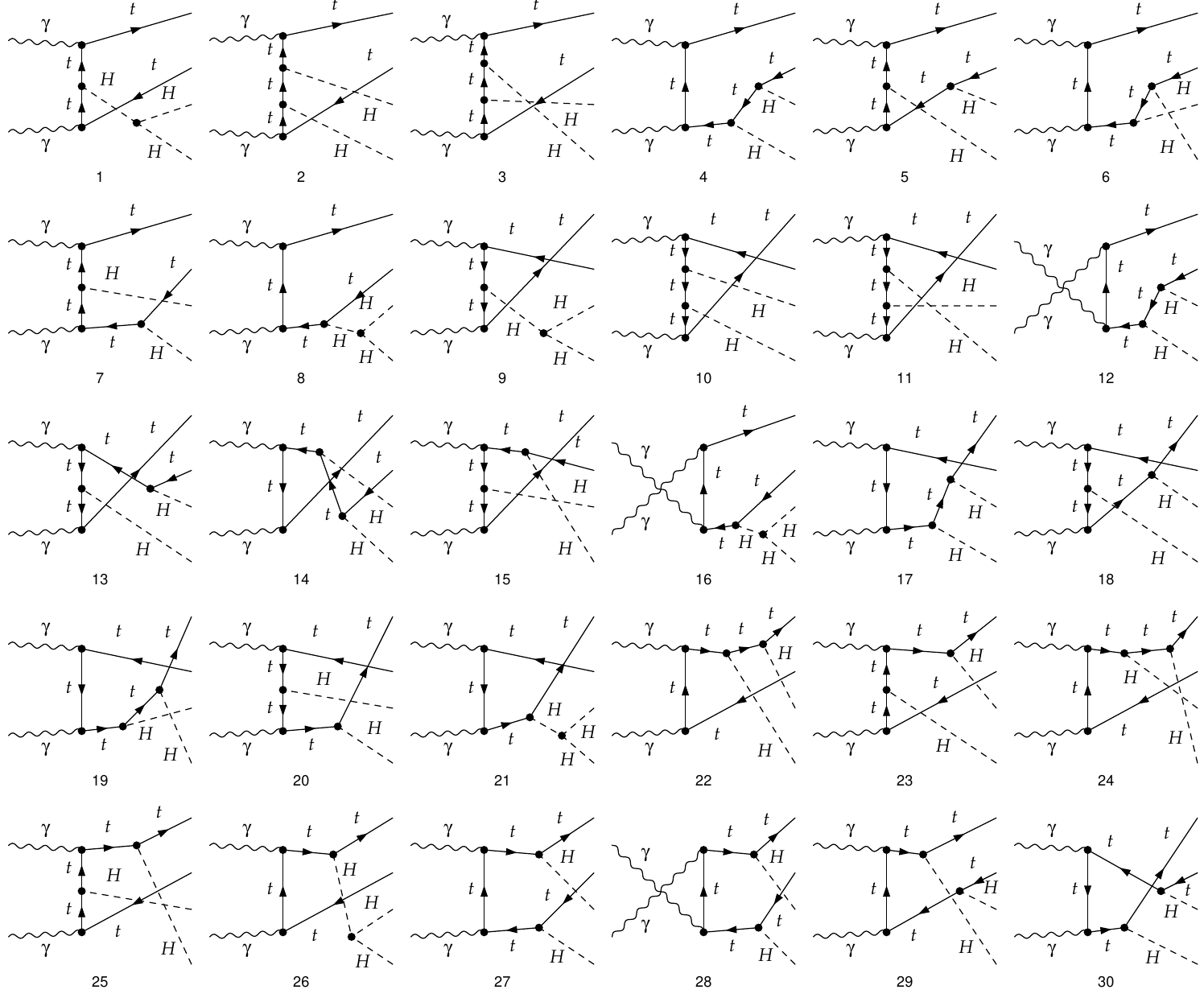}
\caption{Feynman diagrams for the process $\gamma\gamma \to t\bar{t}hh$.}
\label{aa}
\end{figure}
%%%%%%%%%%%%%%%%%%%%%%%%%
At a photon collider, the $\gamma\gamma$ collisions can be also achieved by the inverse Compton scattering of the incident electron- and the laser-beam, the number of events can be calculated by convoluting the cross section of $\gamma\gamma$ collision with the photon beam luminosity distribution:
\begin{eqnarray}
N_{\gamma \gamma \to hh}&=&\int d\sqrt{s_{\gamma\gamma}}
  \frac{d\cal L_{\gamma\gamma}}{d\sqrt{s_{\gamma\gamma}}}
  \hat{\sigma}_{\gamma \gamma \to hh}(s_{\gamma\gamma})
  \equiv{\cal L}_{e^{+}e^{-}}\sigma_{\gamma \gamma \to hh}(s)
\end{eqnarray}
where $d{\cal L}_{\gamma\gamma}$/$d\sqrt{s}_{\gamma\gamma}$ is the photon-beam luminosity distribution. $\sigma_{\gamma \gamma \to hh}(s)$ ( $s$ is the squared center-of-mass energy of $e^{+}e^{-}$ collision) is defined as the effective cross section of $\gamma \gamma \to hh$, which can be written as \cite{photon_collider}
\begin{eqnarray}
\sigma_{\gamma \gamma \to hh}(s)&=&
  \int_{\sqrt{a}}^{x_{max}}2zdz\hat{\sigma}_{\gamma \gamma \to hh}
  (s_{\gamma\gamma}=z^2s) \int_{z^{2/x_{max}}}^{x_{max}}\frac{dx}{x}
 F_{\gamma/e}(x)F_{\gamma/e}(\frac{z^{2}}{x})
\end{eqnarray}
Here $F_{\gamma/e}$ denotes the energy spectrum of the back-scattered photon for the unpolarized initial electron and laser photon beams, which is given by
\begin{eqnarray}
F_{\gamma/e}(x)&=&\frac{1}{D(\xi)}\left[1-x+\frac{1}{1-x}-\frac{4x}{\xi(1-x)}
  +\frac{4x^{2}}{\xi^{2}(1-x)^{2}}\right]
\end{eqnarray}
with
\begin{eqnarray}
D(\xi)&=&(1-\frac{4}{\xi}-\frac{8}{\xi^{2}})\ln(1+\xi)
  +\frac{1}{2}+\frac{8}{\xi}-\frac{1}{2(1+\xi)^{2}}.
\end{eqnarray}
Here $\xi=4E_{e}E_{0}/m_{e}^{2}$ ($E_{e}$ is the incident electron energy, where $E_{0}$ is the initial laser photon energy) and $x=E/E_{0}$ with $E$ being the energy of the scattered photon moving along the initial electron direction. In our calculations, we take the parameters as $\xi=4.8$, $D(\xi)=1.83$ and $x_{max}=0.83$ \cite{photon_collider}.
At photon collider, $\gamma\gamma \to t\bar{t}hh$ process can occur only through the $t$-channel. The corresponding Feynman diagrams for $\gamma\gamma \to t\bar{t}hh$ are shown in Fig.\ref{aa}.

%%Fig.3 %%%%%%%%%%%%%%%%%%%%
\begin{figure}[ht]
\centering
\includegraphics[width=4in,height=3in]{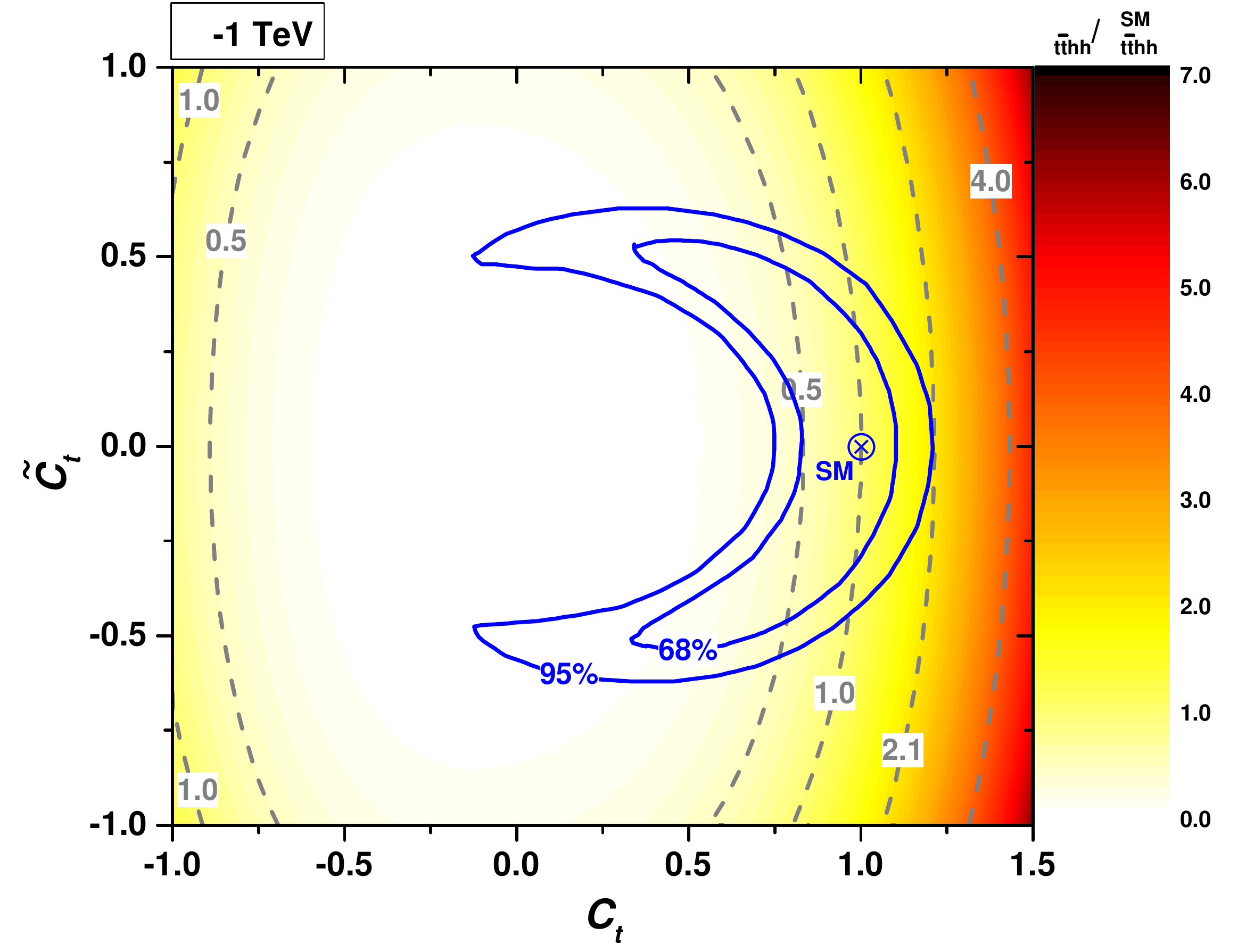}
\caption{Same as Fig.\ref{ratio-hadron}, but ratios of $\sigma^{\gamma\gamma \to t\bar{t}hh}/\sigma^{\gamma\gamma \to t\bar{t}hh}_{SM}$ at a photon collider with $\sqrt{s}=1$ TeV.}
\label{ratio-aa}
\end{figure}
%%%%%%%%%%%%%%%%%%%%%%%%%
In Fig.\ref{ratio-aa}, we present the ratios of $\sigma^{\gamma\gamma \to t\bar{t}hh}/\sigma^{\gamma\gamma \to t\bar{t}hh}_{SM}$ in the plane of $\tilde{c}_t-c_t$ at photon collider with $\sqrt{s}=1$ TeV. Similar to the process $pp \to t\bar{t}hh$, the cross section of $\gamma\gamma \to t\bar{t}hh$ is determined by top-Higgs coupling. From Fig.\ref{ratio-aa}, we can see that the allowed maximal cross section of $\gamma\gamma \to t\bar{t}hh$ can be 2 times as large as the SM prediction, which occurs at $c_t=1.2$ and $\tilde{c}_t=0$. The pseudo-scalar component in $t\bar{t}h$ always reduce the production rate of $\gamma\gamma \to t\bar{t}hh$.

\begin{table}[ht]
\caption{The SM cross sections of signal $t\bar{t}hh$ and the backgrounds $t\bar{t}hb\bar{b}$, $t\bar{t}b\bar{b}b\bar{b}$, $t\bar{t}Zb\bar{b}$ at a $e^+e^-$ collider with $\sqrt{s}=1$ TeV.}
\begin{center}
\begin{tabular}{|c|c|c|c|c|c|}
\hline
$e^+e^-$ &$t\bar{t}hh$  &$t\bar{t}hb\bar{b}$ &$t\bar{t}b\bar{b}b\bar{b}$ &$t\bar{t}Zb\bar{b}$   \\
\hline
$\sigma$ ($10^{-3}$ fb) &$13.55$ &6.69 &5.52  &16.24  \\
\hline
\end{tabular}
\end{center}
\label{ilc}
\end{table}
In Table \ref{ilc}, we present the cross sections of $e^+e^- \to t\bar{t}hh$ and the backgrounds $e^+e^- \to t\bar{t}hb\bar{b}, t\bar{t}b\bar{b}b\bar{b}, t\bar{t}Zb\bar{b}$ for $\sqrt{s}=1$ TeV. It can be seen that the cross section of $t\bar{t}hh$ can reach $13.55\times 10^{-3}$ fb in the SM due to the suppression of phase space. Given the enhancement effect in $t\bar{t}hh$ from the CP-violating top-Higgs couplings, if a $e^+e^-$ collider could deliver an integrated luminosity up to 10 ab$^{-1}$, there will be about hundreds of $t\bar{t}hh$ events, which may be potentially used to test the anomalous top-Higgs couplings. However, the exact detector configurations are not finalized. Therefore, the background estimation and signal extraction strategies would be largely dependent on the detector designs and trigger conditions \cite{beam}, which is beyond the scope of our study. Besides, we find that the cross section of $\gamma\gamma \to t\bar{t}hh$ can only reach $3.74\times10^{-4}$ fb in the SM for $\sqrt{s}=1$ TeV and is hard to be detected at a photon collider.

\section{Conclusion}\label{section4}
As a complementary to the di-Higgs production, $t \bar{t} hh$ process may open a new avenue to measure di-Higgs physics at the LHC and a future 100 TeV $pp$ collider or a high energy $e^+e^-$ collider. In this paper, we studied the impact of CP-violating top-Higgs coupling on $t\bar{t}hh$ production at the LHC, a $e^+e^-$ and a 100 TeV hadron collider under the current Higgs data. Within the allowed parameter region, we find that the cross section of di-Higgs production at the LHC-14 TeV, $e^+e^-$-1 TeV and VHE-LHC/SPPC-100 TeV can be enhanced up to about 2 times the SM predictions since a large deviation of top-Higgs coupling is still allowed. It is expected that the future precise measurement of Higgs coupling will reveal the nature of top-Higgs interaction and improve the sensitivity of the determination of Higgs self-coupling through $t\bar{t}hh$ production.

\acknowledgments
Ning Liu acknowledges Dr Archil Kobakhidze for his warm hospitality in the University of Sydney. This work is supported by the National Natural Science Foundation of China (NNSFC) under grants Nos. 11275057, 11305049 and 11405047, by Specialised Research Fund for the Doctoral Program of Higher Education under Grant No. 20134104120002 and by the Startup Foundation for Doctors of Henan Normal University under contract No.11112, by the China Postdoctoral Science Foundation under Grant No. 2014M561987 and the Joint Funds of the National Natural Science Foundation of China (U1404113).

%%%%%%%%%%%%%%%%%%%%%%%%%%%%
%\clearpage

\bibliographystyle{unsrt}

\end{document}